\documentclass[prd,aps,eqsecnum,floatfix,nofootinbib,preprint,tightenlines]{revtex4}

\usepackage{latexsym}
\usepackage{graphicx}
\usepackage{multirow}
\usepackage{amsmath}
\usepackage{hyperref}  

\def\bibi{\bibitem}

\def\a{\alpha}

\def\c{\chi}
\def\d{\delta}
\def\e{\epsilon}                
\def\g{\gamma}

\def\l{\lambda}
\def\m{\mu}
\def\n{\nu}

\def\p{\pi}                     
\def\s{\sigma}                  
\def\t{\tau}

\def\D{\Delta}

\def\L{\Lambda}

\def\P{\Pi}

\def\S{\Sigma}

\def\cl{{\cal L}}

\def\car{{\cal R}}

\def\ct{{\cal T}}

\def\cbo{{\,\raise-.15ex\Sc [\,}}                       

\def\svev#1{\left\langle #1\right\rangle}       

\def\ddt#1{{\buildrel {\hbox{\LARGE .\kern-2pt.}} \over {#1}}}

\def\ie{\mbox{\it i.e.}}
\def\eg{\mbox{\it e.g.}}

\def\tr{{\rm tr}\,}

\def\half{{1\over 2}}

\def\ttl#1{{\it #1}}

\long\def\symbolfootnote[#1]#2{\begingroup%
\def\thefootnote{\fnsymbol{footnote}}\footnote[#1]{#2}\endgroup}

\long \def \blockcomment #1\endcomment{}

\def\bc{\overline{\c}}
\def\Cw{C_{\rm w}}

\begin{document}

\begin{center}
\vspace{10mm}
{\large\bf
Vacuum alignment and lattice artifacts: Wilson fermions
}
\\[12mm]
Maarten Golterman$^a$
and Yigal Shamir$^b$
\\[8mm]
{\small\it
$^a$Department of Physics and Astronomy\\
San Francisco State University, San Francisco, CA 94132, USA}%
\\[5mm]
{\small\it $^b$Raymond and Beverly Sackler School of Physics and Astronomy\\
Tel Aviv University, Ramat~Aviv,~69978~ISRAEL}%
\\[10mm]
{ABSTRACT}
\\[2mm]
\end{center}

\begin{quotation}
Confinement in asymptotically free gauge theories is accompanied
by the spontaneous breaking of the global flavor symmetry.
If a subgroup of the flavor symmetry group is coupled weakly
to additional gauge fields, the vacuum state tends to align
such that the gauged subgroup is unbroken.
Independently, a lattice discretization of the continuum theory
 typically reduces the manifest flavor symmetry, and, in addition,
can give rise to new lattice-artifact phases with spontaneously broken symmetries.
Here, we study the interplay of these two phenomena for Wilson
fermions, using chiral lagrangian techniques.
We consider two examples: electromagnetic
corrections to QCD, and a prototype
composite-Higgs model.
\end{quotation}

\renewcommand{\thefootnote}{\arabic{footnote}} \setcounter{footnote}{0}

\newpage
\section{\label{Intro} Introduction}
Over the last few years, the computation of certain hadronic quantities using
lattice QCD has become so accurate that electromagnetic effects, while typically
small, need to be included in order to further improve on present errors
\cite{NTlat13}.  A further reduction of the lattice spacing is also needed
in order to suppress competing discretization effects.

There may indeed be a real competition between electromagnetic effects
and lattice artifacts:   Both
can have a non-trivial influence on the phase diagram of the lattice theory.
First, given a strongly interacting gauge theory, let us weakly couple
a subgroup of the flavor symmetry to dynamical gauge fields
(``weak gauge fields,'' for short).
It was observed long ago that the weak gauge fields
can influence the symmetry-breaking pattern.  Their coupling to
unbroken flavor generators tends to stabilize the vacuum, whereas the
coupling to broken generators tend to destabilize it, a
phenomenon usually referred to as ``vacuum alignment'' \cite{MP1980}.
Furthermore, depending on the resulting
alignment, some of the Nambu--Goldstone bosons (NGBs) associated with the
symmetry breaking may acquire a mass, thereby becoming pseudo-NGBs.
An example is the QED-induced
mass splitting between the charged and neutral pions in QCD.

Even without weak gauge fields, a non-trivial phase
structure can also emerge at non-zero lattice spacing.   An example is the possible appearance of a so-called Aoki phase in two-flavor QCD with
Wilson fermions.   Depending on details of the regularization, a phase
can appear in which isospin is spontaneously broken to a $U(1)$
subgroup, alongside with parity \cite{Aoki1983,Creutz1995,ShSi1998}.

It is interesting to study what happens when both effects are at work.
For instance, in lattice QCD with two degenerate Wilson fermions,
what would happen to the Aoki phase if QED is turned on,
or if all the isospin generators are coupled to weak gauge fields?

Similar questions arise beyond the realm of QCD.   The
existence of a light Higgs particle has revived interest in composite
Higgs models, in which a strongly coupled gauge theory breaks its
flavor symmetry dynamically at the TeV scale, producing a massless meson
with the quantum numbers of the Higgs among the NGBs associated
with the breaking of the symmetry.
The flavor currents of this strongly interacting theory can be coupled
to a number of weak gauge fields, with the Standard Model's
electro-weak gauge fields among them.   Electro-weak
symmetry breaking is then arranged to take place through the effective
potential generated for the NGBs of the strongly coupled theory
by the weak dynamics.  A prototype example of such a theory is the
``Littlest Higgs'' model of Ref.~\cite{LSTH}.   In this theory the flavor
symmetry group is $SU(5)$, spontaneously broken by the strong
dynamics to $SO(5)$.   Weak gauge fields are coupled to an
$[SU(2)\times U(1)]^2$ subgroup of $SU(5)$,
with the Standard Model's electro-weak
gauge fields coupling to the diagonal subgroup of $[SU(2)\times U(1)]^2$,
which is also a subgroup of
the unbroken $SO(5)$.

A basic tool used in the phenomenological literature is the non-linear
sigma model describing the (pseudo-) NGBs
(for recent reviews, see Refs.~\cite{RC2010,MP2005}).  Such a low-energy effective
theory requires an ``ultraviolet completion.''  In many cases,
the underlying theory can be taken to be a confining gauge theory,
which, in turn, can be studied on the lattice.  One can then
use numerical methods in order to determine the low-energy constants (LECs)
relevant for electro-weak physics.  Since not only the precise values
of the LECs, but even their signs are usually outside the scope of the non-linear
sigma model, their determination is crucial if we are to confirm
that the correct symmetry-breaking pattern indeed takes place.\footnote{For
realistic studies of the phenomenology of such models, the
top-quark sector should also be taken into account.}   Again, the
question arises whether lattice artifacts might have an effect on the
phase structure, possibly distorting the alignment properties of the continuum
theory.

In this article, we consider these questions in the context of
strongly coupled lattice gauge theories with Wilson fermions.
The use of Wilson fermions means that axial symmetries are explicitly broken
by the discretization, and they are recovered only in the continuum limit.
In the two-flavor theory, this leads to
the practical limitation that weak dynamical gauge fields can be coupled
to isospin generators only, and not to the axial generators.

In order to realize the $SU(5)/SO(5)$ non-linear sigma model
we envisage a confining theory with 5 Weyl fermions in a real representation of
the strong gauge group \cite{MP1980}.
In the continuum, this strongly interacting theory
can equivalently be formulated in terms of Majorana fermions.
Transcribing the latter theory to the lattice is straightforward.
But, once again, if we use Wilson fermions, only the $SO(5)$ flavor symmetry
is preserved, because it is vectorial in the Majorana formulation.
The remaining symmetries (which generate the coset $SU(5)/SO(5)$) are axial.
They are explicitly broken by the Wilson mass term, again to be recovered
only in the continuum limit.  On the lattice we thus consider
only dynamical weak gauge fields for subgroups of $SO(5)$.
As we will see, this is sufficient to gain access to LECs of the
low-energy effective theory that are of interest to phenomenology.

In Sec.~\ref{QCD} we will consider two-flavor QCD with Wilson fermions,
and investigate what happens if we gauge all isospin generators, or if we
gauge only the $U(1)$ subgroup for the $I_3$ component of the photon.
We will consider the lowest-order pion effective potential, containing terms linear in the quark mass, quadratic in the
lattice spacing, and linear in the fine-structure constant, assuming that
these are all of a comparable magnitude.
In Sec.~\ref{SU5} we will then consider the $SU(5)/SO(5)$ non-linear sigma model,
with the weak gauge fields those of the Standard Model group
$SU(2)_L\times U(1)_Y$, in a similar framework.   Because of the more
complicated form of the effective potential, we will not be
able to fully explore the phase diagram that may arise from discretization
effects.  However, a quadratic fluctuation analysis around the vacuum
of the continuum theory will still lead to non-trivial observations.
The final section contains our conclusions,
and a proof of vacuum alignment for the continuum
$SU(5)/SO(5)$ theory is relegated to an appendix.

\section{\label{QCD} Two-flavor QCD with Wilson fermions}
Following Ref.~\cite{ShSi1998}, we start from the effective potential for the
pions of two-flavor lattice QCD with Wilson fermions,\footnote{For reviews
of chiral perturbation theory for QCD with Wilson fermions, see Refs.~\cite{MG2009,StSh2006,OB2004}.}
\begin{eqnarray}
\label{2fpot}
V_{\rm eff}&=&-\frac{c_1}{4}\,\tr(\S+\S^\dagger)+\frac{c_2}{16}\,
\left(\tr(\S+\S^\dagger)\right)^2\\
&=&-c_1\s+c_2\s^2\ ,\nonumber
\end{eqnarray}
in which
\begin{equation}
\label{Sigma}
\S=\s+i\sum_a\t_a\p_a\ ,\qquad \s^2+\sum_a\p_a^2=1 \ ,
\end{equation}
is the non-linear $SU(2)$ matrix built out of the isospin triplet of
pion fields $\p_a$, with $\t_a$ the three Pauli matrices.  The parameter
$c_1$ is linear in the PCAC quark mass $m$, while $c_2$ is proportional to
the square of the lattice spacing $a$.\footnote{Terms linear in the
lattice spacing break the symmetry in exactly the same way as the
term linear in the quark mass, and are thus absorbed into the term
proportional to $c_1$.   Since both $c_1$ and $c_2$ (and $c_3$
in Sec.~\ref{photon} below) have mass
dimension equal to four, appropriate powers of $\L_{\rm QCD}$
will always be understood.}   Higher order terms in the chiral expansion of
$V_{\rm eff}$ will be neglected, since they do not qualitatively affect the
phase diagram (unless at least one of the leading-order terms vanishes).

In the continuum limit, $c_2=0$, and there is a first-order phase transition
when $c_1$, \ie, the quark mass $m$, changes sign: the condensate
$\S_0=\langle\S\rangle$ realigns from $\S_0=+1$ for $c_1>0$
to $\S_0=-1$ for $c_1<0$.
At non-zero lattice spacing, this conclusion does not change if $c_2<0$,
because the $c_2$ term in $V_{\rm eff}$ is minimized
for $\S_0=\pm 1$, irrespective of the sign of $\S_0$.\footnote{In the
large-$N_c$ limit, $c_2<0$ is excluded \cite{sign}, but at finite $N_c$
both signs are possible.} Compared to the continuum theory,
the difference is that for $c_2<0$ the pion masses
do not vanish at the transition; instead, they are all
degenerate, and of order $\sqrt{-c_2}\propto a$.

For $c_2>0$, the minimum of $V_{\rm eff}$ is reached at
\begin{equation}
\label{Vmin}
\langle\s\rangle=
\left\{\begin{array}{ll}
1\ ,&c_1\ge 2c_2\ ,\\
\frac{c_1}{2c_2}\ ,&-2c_2<c_1<2c_2\ ,\\
-1\ ,&c_1\le -2c_2\ .
\end{array}\right.
\end{equation}
For $|c_1|<2c_2$ we find that $|\langle\s\rangle|<1$, which implies that $\langle\p_a\rangle\ne 0$.   $SU(2)$ isospin is spontaneously broken to a $U(1)$
subgroup,\footnote{For the reason that the Vafa--Witten
theorem \cite{VW1983} does not apply inside the Aoki phase, see Ref.~\cite{ShSi1998}.} and two of the three pions become massless as the NGBs associated
with this symmetry breaking.  This region in the phase diagram is the
Aoki phase.
Clearly, in order to probe the Aoki phase
transition,
the couplings $c_1\sim c_2$ have to be of the same magnitude.
We may take the direction of symmetry breaking to
point in the $\t_3$ direction, so that $\p^\pm$ are the NGBs, while $\p^0$
is massive inside the Aoki phase.   At the phase boundaries $|c_1|=2c_2$
all three pions are degenerate and massless, even though $c_1\propto m$
does not vanish.
In the continuum limit, $c_2\propto a^2\to 0$, and the
Aoki phase shrinks to zero; the continuum limit at $c_2=c_1=0$ yields
QCD with two massless quarks.

Inside the Aoki phase of the lattice theory,
parity is spontaneously broken as well.
In the continuum, if we take the vacuum $\svev{\S}=\pm 1$,
parity acts as $\S\to \S^\dagger$.  Since the symmetry is $SU(2)_L\times SU(2)_R$,
any expectation value $\svev{\S}\in SU(2)$ can be rotated
to $\svev{\S}=\pm 1$ using, \eg, an $SU(2)_L$ transformation.  Thus,
if we would want to expand around an equivalent vacuum $\svev{\S}\ne\pm 1$,
parity would merely take a more complicated form.
By contrast, on the lattice the axial symmetries are explicitly broken.
Vacua with different values of $\svev{\s}$ are inequivalent, and,
for any $\svev{\p_a}\ne 0$, parity is broken spontaneously.

\subsection{\label{isospin} Gauging isospin}
We now consider what happens if we gauge isospin, with a gauge coupling
$g$ weak enough that we can analyze the effect on the phase diagram by
considering the order-$g^2$ correction to $V_{\rm eff}$.   We expect that
non-trivial modifications of the scenarios reviewed above
may occur when $g^2\sim c_1\sim c_2$, or,
equivalently, $g^2\sim m/\L_{\rm QCD}\sim (a\L_{\rm QCD})^2$.

In order to find the order-$g^2$ part of $V_{\rm eff}$ we proceed as
follows.   The lowest order chiral effective action contains a term
\begin{equation}
\label{chptaction}
\cl=\frac{f^2}{8}\;\tr\!\left((D_\m\S)^\dagger D_\m\S\right)\ ,
\end{equation}
where $f$ is the pion decay constant in the chiral limit, and
\begin{equation}
\label{covder}
D_\m\S=\partial_\m\S+ig[V_\m,\S]\ ,
\end{equation}
with $V_\m=\sum_a V_{\m,a}\t_a/2$ the isospin gauge field.\footnote{The gauging
of the vector symmetries leads to explicit breaking of the axial symmetries.}
Upon working out the non-derivative part of $\cl$,
\begin{equation}
\label{Lnonder}
\frac{g^2f^2}{4}\;\tr\!\left(V_\m^2-V_\m\S V_\m\S^\dagger\right)\ ,
\end{equation}
we see, first of all, that the weak gauge fields $V_\m$ remain massless
on the isospin-symmetric vacua $\S_0=\pm 1$.   Furthermore, integrating
over the weak gauge fields, we find the leading order contribution to
the effective potential~(\ref{2fpot}):\footnote{The effective potential
due to the weak gauge fields always has a similar form,
even if some of the weakly gauged symmetries are spontaneously broken.
The reason is that the gauge bosons' mass will be proportional to $gf$,
and thus gauge-field mass
effects only show up in the effective potential at order $g^4$.}
\begin{equation}
\label{dVeff}
\D V_{\rm eff}=-\frac{g^2c_3}{8}\, \sum_a\tr\left(\t_a\S\t_a\S^\dagger\right)\ ,
\end{equation}
in which $c_3$ is independent of $g^2$ to leading order.   From Ref.~\cite{EW1983} we know that $c_3>0$.
Using Eq.~(\ref{Sigma}), we find for the effective potential
\begin{equation}
\label{Veffadd}
V_{\rm eff}+\D V_{\rm eff}=-c_1\s+(c_2-g^2c_3)\s^2+\mbox{constant}\ .
\end{equation}
The effect of the weak gauge fields $V_\m$ on the phase diagram is
very simple:  the parameter $c_2$ gets shifted to $c_2-g^2c_3$.
If $c_2<0$, the transition when $c_1$ goes through zero remains
first order.   Even in the continuum limit, when $c_2=0$, all pions
acquire a mass proportional to $\sqrt{g^2c_3}\propto g$.   If $c_2>0$,
the Aoki transition changes into a first-order transition when
the lattice spacing becomes small enough such that $c_2<g^2c_3$.   In other
words, the Aoki phase gets pushed away from the continuum
limit.

\subsection{\label{photon} Coupling the photon}
The situation changes when we restrict the gauge field to
$V_\m=A_\mu Q$, with $Q=\mbox{diag}(2/3,-1/3)
=1/6+\t_3/2$,
and $g=e$, the electric charge, so that $A_\m$ is the photon field.
In that case, the shift in the effective potential becomes
\begin{equation}
\label{dVeffem}
\D V_{\rm eff}^{\rm em}=-\frac{e^2c_3}{8}\; \tr\!\left(\t_3\S\t_3\S^\dagger\right)\ ,
\end{equation}
with the same coefficient $c_3$ as in Eq.~(\ref{dVeff}).
Using Eq.~(\ref{Sigma}) again,
\begin{equation}
\label{Veffem}
V_{\rm eff}+\D V_{\rm eff}^{\rm em}=-c_1\s+c_2\s^2-\frac{e^2c_3}{2}\,(\s^2+\p_3^2)\ .
\end{equation}
Again, the analysis of this effective potential is very simple.   Since $c_3>0$,
minimizing the effective potential requires that $\langle\s\rangle^2+\langle\p_3\rangle^2=1$, \ie,
$\langle\p_1\rangle=\langle\p_2\rangle=0$, irrespective of the values of $c_1$ and $c_2$.  If $c_2<0$,
$\langle\s\rangle=\pm 1$ depending on the sign of $c_1$,
the phase transition is first order, and takes place at $c_1=0$.   The term
proportional to $c_3$ raises the charged pion mass relative to the
neutral pion mass \cite{EMpion}.

If $c_2>0$, and $|c_1|<2c_2$ so that we are in the Aoki phase,
again $\s=c_1/(2c_2)$ as in Eq.~(\ref{Vmin}), and therefore
\begin{equation}
\label{p3vev}
\langle\p_3\rangle=\sqrt{1-\frac{c_1^2}{4c_2^2}}\ .
\end{equation}
Isospin is explicitly broken by the coupling to QED, but parity is
spontaneously broken in the Aoki phase, and there still is a second order
phase transition.   Inside the Aoki phase, the pion masses are
\begin{eqnarray}
\label{pionmasses}
m^2_\pm&=&e^2 c_3 f^{-2}\ ,\\
m^2_0&=&2c_2\left(1-\frac{c_1^2}{4c_2^2}\right) f^{-2}\ .\nonumber
\end{eqnarray}
We see that, depending on the relative size of the parameters
$c_1$, $c_2$ and $e^2c_3$, the neutral pion might even be heavier than the
charged pion, even though in the continuum limit Witten's inequality
implies that this can never be the case \cite{EW1983}.   The reason is
that now we have a competition: electromagnetic effects
increase the charged pion mass relative to the neutral pion mass;
whereas the lattice artifacts that give rise to the
breaking of isospin in the Aoki phase create an opposite effect,
since the charged pions are the NGBs of this symmetry breaking.

\section{\label{SU5} Littlest Higgs}
In this section we present an analysis of the Littlest Higgs model of
Ref.~\cite{LSTH}\footnote{See also Ref.~\cite{MP2005} for a review.}
that parallels what we did for QCD in Sec.~\ref{QCD}.   First,
we very briefly review the necessary ingredients of this theory in the
continuum, in Sec.~\ref{SU5cont}, including the coupling to the Standard
Model gauge fields.  We next consider the Aoki
phase for this theory, without the weak gauge fields, in Sec.~\ref{SU5Aoki}.   In
Sec.~\ref{SU5comb} we then consider the competition between the
effective potential generated by the weak gauge fields and that generated
by lattice artifacts in the determination of the phase diagram.

\subsection{\label{SU5cont} Littlest Higgs -- continuum}
We consider a strongly coupled gauge theory with 5 Weyl fermions
in a real representation of the (unspecified) strong gauge
group.   This theory has an $SU(5)$ flavor symmetry which we assume
to be broken to $SO(5)$ by a bilinear fermion condensate, resulting in 14 NGBs
parametrizing the coset $SU(5)/SO(5)$.   In order to construct the
effective theory for these NGBs, we introduce the non-linear field
\begin{equation}
\label{Sigmasu5}
\S=\exp(i\P/f)\S_0\exp(i\P^T/f)=\exp(2i\P/f)\S_0\ ,
\end{equation}
with $\S_0=\langle\S\rangle$ given by\footnote{Relative to Ref.~\cite{LSTH}
we interchanged the 3rd and 5th rows and columns in the form for
$\S_0$.}
\begin{equation}
\label{vacSU5}
\S_0=\left(\begin{array}{ccccc} 0 & 0 & 1 & 0 & 0\cr
0 & 0 & 0 & 1 & 0\cr
1 & 0 & 0 & 0 & 0\cr
0 & 1 & 0 & 0 & 0\cr
0 & 0 & 0 & 0 & 1\cr
\end{array}\right)\ .
\end{equation}
Since the bilinear fermion condensate is symmetric in its $SU(5)$ indices,
so is $\S$.   Therefore, $\S$ transforms into $U\S U^T$ with $U\in SU(5)$,
and this leads to the form~(\ref{Sigmasu5}) for $\S$ in terms of the ``pion''
field $\P$, which satisfies $\S_0\P^T=\P\S_0$.   The generators $T$ of
the unbroken $SO(5)$ obey the relation $\S_0T^T=-T\S_0$.

The Standard Model $SU(2)_L$ gauge fields $W_{\m a}$ are coupled to an
$SU(2)$ subgroup of $SO(5)$ generated by \cite{LSTH}
\begin{equation}
\label{Wgens}
Q_a=\left(\begin{array}{ccc} \half\t_a & 0 & 0 \cr
0 & -\half\t^T_a & 0 \cr
0 & 0 & 0 \cr
\end{array}\right)\ ,\qquad a=1,\ 2,\ 3\ ,
\end{equation}
where again $\t_a$ are the Pauli matrices.   The $SU(2)$ generated by the
$Q_a$ is an invariant subgroup of the $SO(4)$ group
defined by embedding its elements in the upper-left $4\times 4$
block of the $SO(5)$ matrices.

The leading-order effective potential for the $\S$ field,
obtained by integrating over the $W$ fields, is given by
\begin{equation}
\label{effpotweak}
V_{\rm weak}=g^2 \Cw\,\tr\left(\S Q_a\S^*Q_a^*\right)\ ,
\end{equation}
where a sum over $a$ is implied.  The low-energy constant
$\Cw$ is
analogous to the constant $c_3$ in Eq.~(\ref{dVeff}), and it is
positive, as we show in App.~\ref{proof}, using the relevant result of Ref.~\cite{EW1983}.
In Ref.~\cite{LSTH} more weak gauge fields are coupled to a subgroup of
$SU(5)$ in order to obtain the ``collective'' symmetry breaking typical
of little-Higgs models.   However, the primary goal of a lattice investigation
of this theory would presumably be the determination of the LEC $\Cw$,
which can be probed using any subgroup of $SU(5)$, such as, for instance,
the $SU(2)$ group we introduced in Eq.~(\ref{Wgens}).  As we explain below,
this allows us to maintain all  gauged symmetries (strong and weak)
on the lattice if we choose to work with Wilson fermions.

In Eq.~(\ref{effpotweak}), the minimum value for the trace, $-3$,
is attained for $\S=\S_0$.  Therefore the vacuum is aligned,
\ie, the $W$ fields do not move the vacuum
away from Eq.~(\ref{vacSU5}).   The potential $V_{\rm weak}$ is invariant
under the SO(4) subgroup defined above:  If we transform
$\S\to U\S U^T$ with $U\in SO(4)$, we see that this is equivalent to
keeping $\S$ fixed, while
transforming $Q_a\to U^TQ_a U^*=R_{ab}Q_b$ inside the trace,
with $R$ in the fundamental representation of $SO(3)$.   Here we used that the
$Q_a$ generate an invariant subgroup of $SO(4)$.  Using that
$R_{ab}R_{ac}=(R^T R)_{bc}=\d_{bc}$ the invariance follows.

We may also introduce the hypercharge weak gauge field,
which gauges the $U(1)$ symmetry generated by \cite{LSTH}
\begin{equation}
\label{Y}
Y=\half\,\mbox{diag}\left(1,1,-1,-1,0\right)\ .
\end{equation}
This breaks the $SO(4)$ symmetry explicitly to $SU(2)_L\times U(1)_Y$,
with $SU(2)_L$ the group to which the $W$ fields couple.
The new contribution to the effective potential is
\begin{equation}
\label{VY}
V_Y=g'^2\Cw\,\tr\left(\S Y\S^* Y\right)\ ,
\end{equation}
where the constant $\Cw$ is the same
as in Eq.~(\ref{effpotweak}), and $g'$ the hypercharge gauge coupling.

In order to move to the lattice,
the strongly interacting theory is first reformulated in terms
of Majorana fermions instead of Weyl fermions.
Now, because the fermions transform in a real representation of the
strong gauge group, a gauge-invariant fermion mass term can be
added to the theory, breaking $SU(5)\to SO(5)$ softly.   Going to the lattice
using Wilson fermions, it is then straightforward to augment this
local mass term with a Wilson mass term as well, in order to avoid
species doublers.  The exact flavor symmetry of the lattice theory
is therefore just $SO(5)$, regardless of the fermion mass.
We expect the full $SU(5)$ symmetry to be restored in the continuum
limit, provided that the single-site Majorana mass is tuned appropriately.
These features are, of course, completely analogous to the usual case
of Wilson-Dirac fermions.

On the lattice, before any weak gauge fields are coupled to the flavor
currents and for a large-enough positive quark mass,
the fermion condensate will be proportional to the unit matrix
(see Sec.~\ref{SU5Aoki} below).
Anticipating this, it is convenient to reformulate the (massless) continuum
effective theory such that this is also the case there.
Starting from Eq.~(\ref{vacSU5})
it is straightforward to find an element $U$ of $SU(5)$ such that
\begin{equation}
\label{newvac}
\S'_0=U\S_0 U^T={\bf 1}\ .
\end{equation}
We also have to transform the generators $Q_a$ and
$Y$ to the new basis, defining
\begin{equation}
\label{Qnewbasis}
W_a\equiv UQ_a U^\dagger\ ,\qquad X\equiv UYU^\dagger\ .
\end{equation}
Since $\S_0'$ is proportional to the unit matrix, the $W_a$ and $X$ are anti-symmetric and hermitian, and
thus purely imaginary.   The potential $V_{\rm weak}+V_Y$ can be written as
\begin{equation}
\label{effpotweak2}
V_{\rm weak}+V_Y=-g^2\Cw\,\tr\left(\S W_a\S^*W_a\right)-g'^2\Cw\,
\tr\left(\S X\S^*X\right)\ .
\end{equation}
After adding $V_{\rm weak}+V_Y$ to the effective theory,
the complete vacuum manifold is the $U(1)$ circle generated by
$T=\mbox{diag}(1,1,1,1,-4)$.\footnote{Since we gauge only
the generators $Q_a=Q_1^a+Q_2^a$ and $Y=Y_1+Y_2$ of Ref.~\cite{LSTH}, the Higgs
field components of $\P$ pick up a mass, see Sec.~\ref{SU5comb} below.}

For Majorana (equivalently, Weyl) fermions there are no separate C and P
symmetries, only a CP symmetry.
The role of CP parallels that of parity in the two-flavor
theory of Sec.~\ref{QCD}.  If we expand the non-linear field
around the unit matrix, CP acts on the pion field as $\P\to-\P$.
Since the vacuum manifold contains the unit matrix, it follows
that CP symmetry is unbroken in the continuum theory.

\subsection{\label{SU5Aoki} Littlest Higgs -- lattice artifacts}
In this subsection, we turn off the weak gauge fields, and consider only
the strongly coupled theory on the lattice, using Wilson-Majorana fermions.

The construction of the effective potential representing the effects of a quark mass and
lattice artifacts to order $a^2$ for the $SU(5)/SO(5)$ effective theory is very similar to the construction for the $(SU(2)_L\times SU(2)_R)/SU(2)$ case
reviewed in Sec.~\ref{QCD}.   The only difference is that more invariants proportional to $a^2$ exist, so
that now the effective potential becomes \cite{BRS2003}
\begin{equation}
\label{VAoki}
V_{\rm Aoki}\!=\!-\frac{c_1}{2}\;\tr(\S+\S^\dagger)+\frac{c_2}{4}
\left(\tr(\S+\S^\dagger)\right)^2-\frac{c_3}{4}
\left(\tr(\S-\S^\dagger)\right)^2+\frac{c_4}{2}\,\tr\!\left(\S^2+\S^{\dagger 2}\right),
\end{equation}
in which $c_1$ is proportional to the (subtracted) quark mass, and
$c_{2,3,4}$ are all proportional to $a^2$.\footnote{For $SU(2)$, the
last three terms collapse to the one term in Eq.~(\ref{2fpot}).}
There is no symmetry relating the theory with $c_1>0$ to that with
$c_1<0$, because no non-anomalous transformation exists that relates
the two.
We will therefore mostly limit ourselves to the choice $c_1\ge 0$ in this article.

On our new basis the pion field $\P$ in Eq.~(\ref{Sigmasu5}) is real and symmetric,
and can thus be diagonalized by an $SO(5)$ transformation.   It follows
that in order to find the minimum of $V_{\rm Aoki}$ we may choose $\S$ in Eq.~(\ref{VAoki}) to be diagonal,
\begin{equation}
\label{Sdiag}
\S=\mbox{diag}\left(e^{i\phi_1},e^{i\phi_2},e^{i\phi_3},e^{i\phi_4},e^{i\phi_5}\right)\ ,
\end{equation}
subject to the constraint
\begin{equation}
\label{constraint}
\sum_{i=1}^5 \phi_i=0 \qquad (\mbox{mod $2\p$})\ .
\end{equation}
Substituting this into Eq.~(\ref{VAoki}) yields
\begin{equation}
\label{VA2}
V_{\rm Aoki}=-\sum_i\left(c_1\,\cos{\phi_i}-2c_4\,\cos^2{\phi_i}\right)
+c_2\,\left(\sum_i\cos{\phi_i}\right)^2+c_3\,\left(\sum_i\sin{\phi_i}\right)^2\ .
\end{equation}
This is not easily minimized, so we will begin with simplifying $V_{\rm Aoki}$ by omitting the double-trace terms, \ie, by
setting $c_2=c_3=0$.
Even with only $c_1$ and $c_4$, the minimization of $V_{\rm Aoki}$ will not be
a simple task, because of the constraint~(\ref{constraint}).

For $c_4<0$, the minimum is at $\phi_i=0$, as in the case of Sec.~\ref{QCD}, and the pseudo-NGBs
remain massive in the limit $c_1\to 0$, as long as $c_4\ne 0$;
their mass is proportional to $\sqrt{c_1-4c_4}$.

For $c_4>0$, we will proceed in
several steps.  First we prove that for $c_1>4c_4$ the solution is again
$\phi_i=0$, so that no symmetry is spontaneously broken.   We will
then analyze the case that $c_1=4c_4-2\e$ with $\e>0$ small,
as well as the case that $c_1=\e$ is small.   Since we may take
$c_4$ to set the overall scale of $V_{\rm Aoki}$, we will set $c_4=1$
in most of the rest of this subsection.

The potential $V_{\rm Aoki}$ is extremized if
\begin{equation}
\label{speq}
\sin{\phi_i}\left(c_1-4\cos{\phi_i}\right)=\l\ ,
\end{equation}
where $\l$ is a Lagrange multiplier enforcing the constraint.   First, let us
ignore the constraint, which is equivalent to setting $\l=0$.   Then, for $c_1>4$, Eq.~(\ref{speq}) implies that $\phi_i=0$, if we also demand the solution
to be the minimum of $V_{\rm Aoki}$.   Since this solution satisfies
the constraint~(\ref{constraint}), we have found the solution we are looking for.
Also, since there is only one minimum for $c_1>4$, it follows by continuity
that the same is true at $c_1=4$.   Therefore, if a phase transitions occurs
at $c_1=4$, this phase transition is second order.

Next, we consider $c_1=4-2\e$, with $\e>0$ small.   Since only a continuous
phase transition may take place,
$\phi_i$ will be small as well, and we thus expand the left-hand side of Eq.~(\ref{speq}) to order $\phi_i^3$:
\begin{equation}
\label{speqexp}
\phi_i\left(-\e+\phi_i^2\right)=\l/2\ .
\end{equation}
{}From this, it follows that for any triple $i,j,k$,
if $\phi_i$ is equal to neither $\phi_j$ nor $\phi_k$, then
\begin{equation}
\label{paireq}
\phi_i^2+\phi_i\phi_j+\phi_j^2=\phi_i^2+\phi_i\phi_k+\phi_k^2=\e\ .
\end{equation}
It follows that either $\phi_k=\phi_j$, or $\phi_k=-\phi_i-\phi_j$.
This provides us with a finite list of options to check, and we find that
$V_{\rm Aoki}$ is minimized for
\begin{subequations}
\label{minnear4}
\begin{eqnarray}
\S=\S_0(4-2\e)&=&\exp\left[i\,\mbox{diag}\left(\phi,\phi,\phi,-3\phi/2,-3\phi/2\right)
\right]\ ,\label{minnear4a}\\
\phi^2&=&\frac{9}{7}\,\e\ .\label{minnearb}
\end{eqnarray}
\end{subequations}
Indeed, a second order phase transition takes place at $c_1=4$, with,
below that value, a  symmetry-breaking pattern
 $SO(5)\to SO(3)\times SO(2)$.   In addition, CP symmetry,
 $\S\to\S^*$, is spontaneously broken as well.   We note that the
 solution~(\ref{minnear4}) cannot be rotated to $\S_0={\bf 1}$,
 because on the lattice the $SU(5)$ transformation that would
 do this is not a symmetry.

We now turn to the case that $c_1=0$.   If $\phi_0$ is a solution of
Eq.~(\ref{speq}), \ie, $\sin{2\phi_0}=-\l/2$, then all possible solutions are
\begin{equation}
\label{c20sol}
\phi_i=\phi_0\ ,\qquad\phi_i=\pi/2-\phi_0\ ,\qquad\phi_i=\pi+\phi_0\ ,\qquad
\phi_i=3\pi/2-\phi_0\ .
\end{equation}
Going through all possibilities for choosing the $\phi_i$, $i=1,\dots,5$
from this list, and demanding that any such choice satisfies the constraint~(\ref{constraint}), yields three degenerate minima for $c_1=0$:
\begin{subequations}
\label{minat0}
\begin{eqnarray}
\S=\S^{(1)}_0(0)&=&\exp\left[(2\p i/5)\,\mbox{diag}\left(1,1,1,1,1\right)\right]\ ,
\label{minat0a}\\
\S=\S^{(2)}_0(0)&=&\exp\left[(2\p i/5)\,\mbox{diag}\left(1,1,1,-3/2,-3/2\right)\right]\ ,
\label{minat0b}\\
\S=\S^{(3)}_0(0)&=&\exp\left[(2\p i/5)\,\mbox{diag}\left(1,-3/2,-3/2,-3/2,-3/2\right)\right]\ .\label{minat0c}
\end{eqnarray}
\end{subequations}
Next, let us consider small $c_1=\e$.  Once again, since the three global
minima at $c_1=0$ are discrete, this can at most lead
to a small shift $\d\phi_i$ away from $2\pi/5$ or $-3\pi/5$ for each $i$.
Expanding $V_{\rm Aoki}$, we find
\begin{subequations}
\label{epsexp}
\begin{eqnarray}
V^{(1)}_{\rm Aoki}&=&\frac{5}{4}\left(3-\sqrt{5}-\e\,\left(\sqrt{5}-1\right)\right)
+\frac{1}{2}\left(1+\sqrt{5}\right)\sum_i\d\phi_i^2\ ,\label{epsexpa}\\
V^{(2)}_{\rm Aoki}&=&\frac{5}{4}\left(3-\sqrt{5}-\frac{\e}{5}\,\left(\sqrt{5}-1\right)\right)
+\frac{1}{2}\left(1+\sqrt{5}\right)\sum_i\d\phi_i^2\ ,\label{epsexpb}\\
V^{(3)}_{\rm Aoki}&=&\frac{5}{4}\left(3-\sqrt{5}+\frac{3\e}{5}\,\left(\sqrt{5}-1\right)\right)
+\frac{1}{2}\left(1+\sqrt{5}\right)\sum_i\d\phi_i^2\ ,\label{epsexpc}
\end{eqnarray}
\end{subequations}
where the superscript on $V_{\rm Aoki}$ refers to which solution in
Eq.~(\ref{minat0}) we are expanding around.
We have expanded to quadratic order in $\d\phi_i$,
dropping terms of order $\e\,\d\phi_i^2$, and we have used that $\sum_i\d\phi_i=0$ because of Eq.~(\ref{constraint}).
For small $c_1=\e>0$ the first minimum, $\S^{(1)}_0(0)$, is the absolute
minimum, and, since the coefficients of the $\d\phi_i^2$ terms
in Eq.~(\ref{epsexp}) are always
positive, the minimum stays at $\S_0(c_1>0)=\S^{(1)}_0(0)$.
The vacuum $\S^{(1)}_0(0)$ only breaks CP.   For $c_1<0$, the solution
$\S^{(3)}_0(0)$ becomes the absolute minimum,
and a first-order transition takes place at $c_1=0$.  This solution breaks $SO(5)$ to $SO(4)$, giving rise to 4 NGBs.

The picture that arises is that there is an Aoki-like phase when $c_4>0$.
For $c_1$ just below $4c_4$, $SO(5)$ breaks to $SO(3)\times SO(2)$,
and there are 6 exact NGBs.  CP is broken as well.
We do not know what happens when
$c_1$ is further decreased, but when we reach $c_1=0$ (while keeping
$c_1>0$) only CP remains spontaneously broken.

We end this section with a few observations on what happens if $c_2$ and
$c_3$ are turned back on.   First, with $c_3=0$, Eq.~(\ref{speq}) becomes
\begin{equation}
\label{speqc2}
\sin{\phi_i}\left(c_1-4\cos{\phi_i}-2c_2\sum_i\cos{\phi_i}\right)=\l\ .
\end{equation}
Our previous symmetric solution, $\phi_i=0$, is still the only solution when $c_1>4c_4+10c_2$.  Moreover, this remains true when
$c_3>0$ as well.

Next, we consider in more detail what happens for smaller values of
$c_1$ when $c_2$ and $c_3$ do not vanish, but are small.   Near $c_1=0$,
the potential remains equal to a constant plus a positive-definite
quadratic form in $\d\phi_i$.  Denoting the new constant piece as
$\d V^{(i)}_{\rm Aoki}$ we find
\begin{equation}
\d V^{(i)}_{\rm Aoki} =
b^{(i)} \left( c_2\,\frac{3-\sqrt{5}}{8} + c_3\,\frac{5+\sqrt{5}}{8}\right) \ ,
\label{c23}
\end{equation}
where $b^{(1)}=25$, $b^{(2)}=1$, and $b^{(3)}=9$.  Since $\d V^{(2)}_{\rm Aoki}$
is smaller than the other two, there will be regions where each
of the solutions, now including $\S^{(2)}_0(0)$, is the global minimum.

For $c_1$ near $4c_4$, we consider the case that
$c_2\sim c_3\sim\e$.   Expanding Eq.~(\ref{speqc2}), we find the $c_2\ne 0$
version of Eq.~(\ref{speqexp}),
\begin{equation}
\label{speqexpc2}
\phi_i\left(-(\e+5c_2)+\phi_i^2\right)=\l/2\ .
\end{equation}
The $c_3$ term does not contribute to this order, because of the
constraint~(\ref{constraint}).    The phase transition now takes place
when $\e+5c_2$ becomes positive, or, equivalently, when
$c_1$ becomes smaller than $4+10c_2$.   Restoring $c_4$, the phase boundary gets shifted from $c_1=4c_4$ to $c_1=4c_4+10c_2$,
consistent with what we already found above.

\subsection{\label{SU5comb} Combined phase diagram}
We now combine the potentials $V_{\rm Aoki}$ of Eq.~(\ref{VAoki}) and
$V_{\rm weak}$ of Eq.~(\ref{effpotweak}).\footnote{We set the hypercharge
gauge coupling $g'=0$, since not much changes in our analysis when
it is turned on.}   The combined potential is invariant under $SO(4)$,
embedded in the upper-left $4\times 4$ block of the $SO(5)$ matrices.
The pion field $\P$, which transforms as the traceless, two-index
symmetric representation of $SO(5)$, decomposes into fields
transforming as the traceless, two-index symmetric representation of $SO(4)$
which we will denote by ${\bf 9}$,
the fundamental representation, denoted by ${\bf 4}$,
and a singlet, denoted by ${\bf 1}$.

A first observation is that, when $c_3\ge 0$, the phase transition
boundary is still at $c_1=4c_4+10c_2$.   This is because for
$c_1\ge 4c_4+10c_2$, the minimum of both $V_{\rm Aoki}$
and $V_{\rm weak}$ is at $\phi_i=0$.
Then, substituting Eq.~(\ref{Sigmasu5}) into the sum of
Eqs.~(\ref{VAoki}) and~(\ref{effpotweak}), we find for the masses
of each of these representations\footnote{In the model of Ref.~\cite{LSTH},
thanks to the presence of more weak gauge fields,
$M_{\bf 4}=0$ in the continuum, allowing its identification
with the Higgs field.}
\begin{eqnarray}
\label{masses}
M_{\bf 1}^2&=&(4/f^2)(c_1-4c_4-10c_2)\ ,\\
M_{\bf 4}^2&=&(4/f^2)(c_1-4c_4-10c_2+3g^2\Cw/4)\ ,\nonumber\\
M_{\bf 9}^2&=&(4/f^2)(c_1-4c_4-10c_2+2g^2\Cw)\ .\nonumber
\end{eqnarray}

We see that indeed $V_{\rm weak}$ leads to mass
splittings consistent with the fact that it breaks $SO(5)$ to $SO(4)$.
For $g=0$, we recover the observation that one enters an Aoki
phase when $c_1-4c_4-10c_2$ becomes negative, even though
the masses alone are not sufficient to deduce the symmetry-breaking
pattern found in Sec.~\ref{SU5Aoki}.   Furthermore, in the continuum limit
($c_2=c_3=c_4=0$) and for vanishing quark mass ($c_1=0$), Eq.~(\ref{masses})
confirms that $V_{\rm weak}$ stabilizes the vacuum manifold
that we have inferred from Eq.~(\ref{effpotweak2}).

An important practical issue facing the lattice simulation of a UV completion
of
this model is how to tune the bare mass towards its critical value
where the (Majorana) fermions become massless.
Starting from large positive $c_1$, in the absence of weak gauge fields
the fermion mass, as well as the masses of all pions, will vanish simultaneously
when $c_1$ reaches $4c_4+10c_2$.
Equation~(\ref{masses}) tells us that, when the weak
gauge fields are dynamical, this is no longer true.  Clearly, the
massless limit of the continuum theory corresponds to vanishing $M_{\bf 1}^2$,
and therefore it is the singlet that must be tuned to criticality in
a lattice simulation.
By contrast, if one were to tune $M_{\bf 4}^2$ or $M_{\bf 9}^2$
to zero in the lattice theory, the curvature in the singlet direction
would have become negative at the origin, implying that
the $SO(4)$ singlet field, $\eta$, has acquired a non-vanishing expectation value.
Since this vacuum is still invariant under the $SO(4)$ symmetry of the
full potential $V$, there are no NGBs.  However,
CP symmetry is spontaneously broken, because for $\langle\eta\rangle\ne 0$
one has $\langle\S\rangle\ne\langle\S^*\rangle$ on the lattice.

We have not been able to minimize the full potential inside
the Aoki phase.  It appears likely that, as one moves towards more
negative values of $c_1-4c_4-10c_2$, the $SO(4)$ symmetry will break
spontaneously, giving rise to some NGBs.

Imagine starting at some small but fixed $c_1-4c_4-10c_2<0$,
and gradually turning on $g$.  For $g=0$,
we have found that the vacuum is given by Eq.~(\ref{minnear4}),
with symmetry breaking $SO(5)\to SO(3)\times SO(2)$.
For $g\ne 0$ the symmetry of the theory is reduced to $SO(4)$.
As long as $g$ is small enough, we expect that the vacuum~(\ref{minnear4})
will be modified by continuous $O(g^2)$ corrections.
One may speculate on how the $SO(3)\times SO(2)$ and $SO(4)$
subgroups of $SO(5)$ align relative to each other.   One possibility
is that the $SO(3)$ is a subgroup of the $SO(4)$, with the spontaneous
symmetry breaking pattern  $SO(4)\to SO(3)$.   Instead of 6, there will
only be 3 exact NGBs.   However, one can verify that in that case
all the $SU(2)$ generators in Eq.~(\ref{Qnewbasis}) are broken,
and these 3 NGBs are thus eaten by the $W_{\m a}$ gauge fields.
Another possibility is that the $SO(2)$ is a subgroup of the $SO(4)$,
while the $SO(3)$ is explicitly broken to another $SO(2)$.
The spontaneous symmetry breaking pattern is now
$SO(4)\to SO(2)\times SO(2)$, yielding 4 NGBs.   In this case,
only two out of three $SU(2)$ generators in Eq.~(\ref{Qnewbasis}) are broken. One
$W$ field stays massless, with 2 exact NGBs remaining in the
spectrum.   According to Ref.~\cite{MP1980}, this scenario would be favored.

We see that, deeper
inside the Aoki phase, it is quite likely that one would encounter
long-range effects mediated by exact NGBs.
The existence of such exact NGBs
is purely a lattice artifact.

\section{\label{concl} Conclusions}
In an asymptotically free gauge theory with massless fermions
one can consider
a number of small perturbations.   In the continuum, one can give
the (Dirac or Majorana) fermions a mass.
One can also couple the fermions to another
dynamical gauge field gauging some of the flavor symmetries,
such that, at the scale where the original gauge theory confines,
the new gauge coupling is weak.
In general, such perturbations break the flavor symmetry
of the strong gauge theory explicitly.

In addition, in order to study such a theory
non-perturbatively, one needs to consider the lattice discretization.
Again, the lattice formulation usually breaks explicitly some of
the flavor symmetries.  Finally, the strong dynamics typically
gives rise to spontaneous symmetry breaking.

In this article we investigated flavor symmetry breaking using
effective field theory techniques in two examples, using Wilson
fermions for the lattice formulation of the theory.   In both cases,
we allowed the weak gauge fields to couple only to a subgroup
of the lattice flavor symmetry group, since otherwise we would need
to consider a chiral gauge theory on the lattice.\footnote{The
definition of chiral gauge theories on the lattice is as yet not a fully
solved problem, see for example Refs.~\cite{LChGT}.}
In many applications
to physics beyond the Standard Model, weak gauge fields coupling
to broken, or axial, generators are also needed.
Nevertheless, the restriction to weak gauge fields
coupled to conserved lattice currents is not a severe limitation,
as it already gives access to LECs whose values are phenomenologically
interesting.  The reason is that, thanks to its symmetry structure,
the (continuum) effective theory is typically characterized
by a very small number of LECs, which are common to
weak gauge fields coupled to both vector and axial generators.

The two examples we considered are QCD with two light
flavors where also (part of) the isospin symmetry group is gauged,
and the Littlest Higgs model of Ref.~\cite{LSTH}.  In the latter case,
only the Standard-Model subgroup of the flavor symmetry group was gauged,
because, among the weak gauge fields of Ref.~\cite{LSTH},
only the electro-weak fields couple to vector currents
of the strongly interacting theory.
Since a lattice gauge theory with Wilson fermions gives rise to a
non-trivial phase diagram at non-zero lattice spacing, this phase
structure can ``interfere'' with the expected effects of the continuum
perturbations from the fermion masses and weak gauge fields.

In the QCD case, in the continuum limit, gauging isospin
leads to all pions acquiring a mass.   However, if lattice spacing effects,
represented in the effective theory through the LEC $c_2$ in Sec.~\ref{QCD},
are large enough, one finds that some of the pions may remain massless
as a pure lattice artifact.   This happens if the LEC $c_2>0$ so that an Aoki phase
exists near the continuum limit.   Moreover, in that case also parity is
spontaneously broken.    In the case that only a $U(1)$ subgroup of
isospin is gauged, all pions can remain massive even for vanishing quark mass, with the neutral
pion mass of order the lattice spacing, but parity can still be spontaneously
broken.  Perhaps surprisingly, the neutral pion can be heavier than the charged pion.

Very similar conclusions are obtained in the case of the Littlest Higgs model,
which we studied in Sec.~\ref{SU5}.   In the continuum, the weak gauge fields
make most of the Nambu--Goldstone bosons of the strong gauge theory
massive, but inside the Aoki phase of the lattice version of the theory,
some of these mesons may again become massless, as a consequence
of lattice artifacts.   Moreover, inside the Aoki phase, CP is spontaneously
broken as well.   Because of the complicated structure of the effective
potential in this case an exhaustive study of the phase diagram is more
difficult, but the message is essentially the same as in the case of QCD
with two flavors.

Our results lead us to the following conjecture.  If a general
subgroup of the unbroken flavor symmetries is gauged, we expect the
boundary
of the Aoki phase to stay at the same location, but the symmetry breaking
pattern inside the Aoki phase can change.  If, however, an invariant
subgroup of the unbroken flavor symmetry group is gauged, the
potential will retain the same flavor symmetry, and, as a result, the
boundary of the Aoki phase itself will shift its location. This
includes the case where the full
unbroken flavor symmetry group is gauged, as in Sec.~\ref{isospin}.\footnote{An example
of a non-trivial invariant subgroup would be an SU(4)/SO(4) non-linear sigma
model, where the lattice flavor symmetry is SO(4), and
in which the invariant subgroup of Sec.~\ref{SU5cont} is gauged.}

Clearly, these results have practical consequences for the lattice study
of electromagnetic effects in hadronic physics and for composite Higgs
models.    The interplay between all three sources of symmetry breaking
(weak gauge fields, fermion masses, and lattice artifacts) will have
to be considered very carefully in order to arrive at valid conclusions
about the continuum limit.  For example, in the context of the
Littlest Higgs model of Sec.~\ref{SU5}, our analysis clarifies how to tune
to the massless limit on the lattice.
It should be straightforward to extend the
analysis framework we developed in this article to gauge theories with
different flavor symmetry groups.

Finally, our conclusions are not limited to lattice
gauge theories with Wilson fermions.  In a companion paper \cite{GS2014}
we find that similar considerations apply to
the use of staggered fermions as well, since staggered fermions also break
continuum flavor symmetries,\footnote{For reviews, see Refs.~\cite{MG2009,stagrev} and
references therein.} and a non-trivial phase structure is
possible in that case as well \cite{AW2004}.  In addition, the same
continuum mass matrix can arise from inequivalent choices of the staggered
mass terms on the lattice, and this can also give rise to a competition
with the effects of the weak gauge fields.

\vspace{3ex}
\noindent {\bf Acknowledgments}
\vspace{3ex}

We thank Santiago Peris for discussions.
MG thanks the School of Physics and Astronomy of Tel Aviv University
and YS thanks the Department of Physics and Astronomy of San Francisco
State University for hospitality.
MG is supported in part by the US Department of Energy, and
YS is supported by the Israel Science Foundation under grants no.~423/09 and~449/13.

\appendix
\section{\label{proof} Proof that $\Cw$  is positive}
In this appendix we prove that $\Cw$ in Eq.~(\ref{effpotweak}) is positive.
The structure of the proof is similar to the proof that the
electromagnetic contribution to $m_{\p^\pm}^2-m_{\p^0}^2$ in QCD is
positive \cite{EW1983}.

We will consider the case of a strongly coupled gauge theory with
$N_f$ Weyl fermions in a real representation of the strong
gauge group.   This theory has an $SU(N_f)$ flavor symmetry,
which is spontaneously broken to $SO(N_f)$.   For the purpose of this
appendix, it is convenient to assemble each two-component Weyl fermion
and its anti-fermion field into a four-component
Majorana fermion $\c_i$, $i=1,\dots,N_f$.
The continuum action is then $\half \bc_i \g_\m D_\m \c_i$, where
$D_\m$ is the covariant derivative in the real representation.
Here $\bc_i=\c_i^TC\car$ by definition,
where $C$ is the usual charge-conjugation matrix, and $\car$ is a
matrix such that $\car^T=\car^\dagger=\car^{-1}=\car$ and $\car \ct_\a\car=-\ct_\a^T$ for the generators $\ct_\a$ of the strong gauge group in the real representation.
The fermion condensate $\svev{\bc_i \c_j}$ is symmetric in the indices
$i,j$, and we will assume that $\svev{\bc_i \c_j}\propto \d_{ij}$.
Thus the unbroken $SO(N_f)$ generators are anti-symmetric, $T_a^T=-T_a$, and the
broken generators for the coset $SU(N_f)/SO(N_f)$ are symmetric, $T_a^T=T_a$.

We introduce an $SU(N_f)$ global spurion $Q=Q_a T_a$,
with $T_a$ the hermitian generators of $SU(N_f)$.\footnote{For the QCD
plus QED case, see Refs.~\cite{RC2010,GS2014,SP2002}.}
The spurion transforms as $Q \to U Q U^\dagger$ for $U\in SU(N_f)$.
The microscopic partition function is%
\footnote{The factor of $i$ in Eq.~(\ref{NfHb})
is erroneously missing in the published version.}
\begin{eqnarray}
  Z(Q) &=&
 \int d[A] d[W] d[\c] \,\exp[-S_{\rm S}(A_\m,\c_i)-S_{\rm W}(W_\m,\c_i,Q)] \,,
\hspace{6ex}
\label{NfHa}\\
  S_{\rm W}(W_\m,\c_i,Q) &=&  \frac{1}{4} F_{\m\n}^2 + igW_\m Q_a J_{\m a} \,,
\label{NfHb}\\
  J_{\m a} &=& \bc_i \g_\m P_R T_{aij} \c_j
  \ = \ \bc_i \g_\m P_L (-T^T)_{aij} \c_j \,.
\label{NfHc}
\end{eqnarray}
Here $A_\m$ is the strong gauge field, and $S_{\rm S}$ the action for the
strong dynamics.
The field $W_\m$ is the weak gauge field, with $F_{\m\n}$ its field strength.
Since we work to order $g^2$, a single gauge field $W_\m$ will be sufficient.  Correspondingly, we may neglect the non-linear part of $F_{\m\n}$.
In this framework, the global $SU(N_f)$ transformations are carried
by the spurions $Q_a$, whereas the field $W_\m$ is invariant.\footnote{It is
also possible to promote the flavor symmetry to a local symmetry
(at least classically), by introducing the gauge fields $W_{\m a} T_a$,
see Ref.~\cite{RC2010}.}

The leading-order effective potential, bilinear in $Q$, is now
\begin{equation}
 V_{\rm eff}  = g^2C_0\, \tr(Q^2) + g^2\Cw\,\tr(Q \S Q^* \S^*) \,,
\label{ChNfCH}
\end{equation}
in which $C_0$ is another constant.  The chiral field $\S$
is a unitary and symmetric matrix, and transforms as $\S\to U\S U^T$.
Using $Q^*=Q^T$,
on the vacuum $\S_0={\bf 1}$ this expression collapses to
\begin{equation}
 V_{\rm vac}  = g^2C_0\, \tr(Q^2) + g^2\Cw\,\tr(Q Q^T) \,.
\label{LvacUO}
\end{equation}
Introducing general linear combinations $Q^V$ and $Q^A$ of
the unbroken and broken $SU(N_f)$ generators,
\begin{equation}
\label{QO}\\
  Q^V = \sum_{T_a=-T_a^T} Q^V_a T_a \,,\qquad
  Q^A = \sum_{T_a=+T_a^T} Q^A_a T_a \,.
\end{equation}
and using that $Q=Q^V+Q^A$
and $Q^T=-Q^V+Q^A$, we may write $V_{\rm vac}$ as
\begin{equation}
 V_{\rm vac}  = g^2C_0\,\tr(Q^V Q^V + Q^A Q^A) - g^2\Cw\, \tr(Q^V Q^V - Q^A Q^A) \,.
\label{LvacVAUO}
\end{equation}
Differentiating twice yields the linear combinations
\begin{subequations}
\label{HgetC}
\begin{eqnarray}
  \frac{\partial}{\partial Q^V_a}
  \frac{\partial}{\partial Q^V_b}\, V_{\rm vac}
  &=& g^2\d_{ab} (C_0-\Cw) \,,
\label{HgetCXX}\\
  \frac{\partial}{\partial Q^A_a}
  \frac{\partial}{\partial Q^A_b}\, V_{\rm vac}
  &=& g^2\d_{ab} (C_0+\Cw) \,.
\label{HgetCLR}
\end{eqnarray}
\end{subequations}
from which we may extract $C_0$ and $\Cw$ separately.

In the microscopic theory
\begin{subequations}
\label{micJJ}
\begin{eqnarray}
  \svev{J_{\m a}(x) J_{\n b}(0)}
  &=& -\tr \svev{\g_\m T_a P_R [\c(x) \bc(0)] \g_\n T_b P_R
      [\c(0) \bc(x)] }
\label{micJJa}\\
  && +\tr \svev{\g_\m T_a P_R [\c(x) \bc(0)] \g_\n T_b^T P_L
      [\c(0) \bc(x)] }
\hspace{5ex}
\nonumber\\
  &=& -\tr(T_a T_b)\, \tr \svev{\g_\m P_R [\c(x) \bc(0)] \g_\n P_R
      [\c(0) \bc(x)] }
\label{micJJb}\\
  && +\tr(T_a  T_b^T) \,\tr \svev{\g_\m  P_R [\c(x) \bc(0)] \g_\n P_L
      [\c(0) \bc(x)] } \,.
\nonumber
\end{eqnarray}
\end{subequations}
where $[\c(x) \bc(y)]$ is the Majorana fermion propagator.
In Eq.~(\ref{micJJb}), in each term the first trace is over flavor indices,
and the second over Dirac and strong gauge-group indices.

The reason for the two terms on the right-hand side is that, with $\c_i$ being
Majorana, two different contractions contribute.
 In the first term on the right-hand side of Eq.~(\ref{micJJ}),
we express both currents using the first expression on the right-hand side of Eq.~(\ref{NfHc}),
and then contract the fermion fields cyclically.  The second term
is obtained by first rewriting $J_{\n b}(0)$ using the second expression on the right-hand side of
Eq.~(\ref{NfHc}), before  cyclically contracting the fermions.

Unlike in the case of QCD, the same two-current correlation function now has both symmetry-preserving
and symmetry-breaking parts.  But these two parts have a
different flavor structure.  Indeed, the flavor structure of the two terms
in Eq.~(\ref{micJJb}) reproduces that obtained at the effective potential
level~(\ref{LvacUO}).  Defining form factors from the contractions
($P^\perp_{\m\n}$ is the transverse projector)
\begin{subequations}
\label{Hform}
\begin{eqnarray}
  q^2 P^\perp_{\m\n}\, \P_0(q^2)
  &=& -\int d^4x\, e^{iqx} \tr \svev{\g_\m P_R [\c(x) \bc(0)] \g_\n P_R
      [\c(0) \bc(x)] } \,, \hspace{5ex}
\label{Hforma}\\
  q^2 P^\perp_{\m\n}\, \P_{\rm w}(q^2)
  &=& \int d^4x\, e^{iqx} \tr \svev{\g_\m  P_R [\c(x) \bc(0)] \g_\n P_L
      [\c(0) \bc(x)] } \,,
\label{Hformb}
\end{eqnarray}
\end{subequations}
one finds that%
\footnote{The factors of 3 in Eq.~(\ref{Cs}) come from tracing over the
transverse projector.  They are erroneously missing in the published version.}
\begin{subequations}
\label{Cs}
\begin{eqnarray}
C_0&=&\frac{3}{16\p^2}\int_0^\infty dq^2 q^2 \P_0(q^2)\,,\label{Csa}\\
\Cw&=&\frac{3}{16\p^2}\int_0^\infty dq^2 q^2 \P_{\rm w}(q^2)\,.\label{Csb}
\end{eqnarray}
\end{subequations}
Finally, we observe that the Dirac structure in Eq.~(\ref{Hformb}) is
identical to that of $\P_{LR}(q^2)$ in QCD, and therefore the proof
in Ref.~\cite{EW1983} that $\P_{LR}(q^2)\ge 0$ applies to $\P_{\rm w}(q^2)$
as well, with the consequence that $\Cw>0$.   We note that the first
of these two integrals is UV divergent, but the second, being an order
parameter, is finite.

\vspace{5ex}

\end{document}